\documentclass[aps,prl,twocolumn,groupedaddress,showpacs,floatfix]{revtex4}
\usepackage{epsfig}
\usepackage{epsf} 
\begin{document}
%

\title{Phonons in the $\beta$-tin, {\it Imma}, and sh phases of
  Silicon from {\it ab initio} calculations} 

\author{Katalin Ga\'al-Nagy}
\altaffiliation{present adress: Dipartimento di Fisica and INFM, 
Universit\`a degli Studi di Milano, via Celoria 16, I-20133 Milano, Italy}
\email{katalin.gaal-nagy@physik.uni-regensburg.de}
\author{Dieter Strauch}
\affiliation{Institut f{\"u}r theoretische Physik, Universit{\"a}t
  Regensburg, D-93040 Regensburg, Germany}

\date{\today}
%
\begin{abstract}
  We present a new interpretation of measured Raman frequencies of a
  high-pressure structure of Silicon which was assigned previously
  to the $\beta$-tin phase. Our results show that the
  $\beta$-tin$\to${\it Imma}$\to$sh phase transitions have been
  already indicated in this experiment which was performed before
  the discovery of the {\it Imma} phase. We have calculated
  phonon-dispersion curves for the $\beta$-tin, {\it Imma}, and sh
  phases of silicon using the plane-wave pseudopotential approach to
  the density-functional theory and the density-functional
  perturbation theory within the local density approximation. With
  the new assignment, the calculated phonon frequencies display an
  excellent agreement with the experimental data, and can be also
  used to determine precisely the transition pressure for the {\it
  Imma}$\to${$\beta$-tin} phase transition. The sh$\to${\it Imma}
  transition is accompanied by soft modes.
\end{abstract}
\maketitle

%
%
In 1993 a new high-pressure phase was found between the $\beta$-tin
(SiII, body centered tetragonal structure) and the sh (SiV, simple
hexagonal structure) phase of silicon \cite{McM93}. This phase was
called {\it Imma} phase after its space group and has a
body-centered orthorhombic structure (SiXI). In the diffraction
experiment, the phase transitions to and from the {\it Imma} phase
appear as of first order with a discontinuity in the volume and the
lattice parameters \cite{McM94}. Previous ab initio calculations of
the phase transitions $\beta$-tin$\to${\it Imma}$\to$sh have
indicated both phase transitions to be of second order \cite{Lew93a}
or both of first order \cite{Chr99}. A theoretical investigation of
elastic stabilities has shown a second-order phase transition
$\beta$-tin$\to${\it Imma} \cite{Heb01}. From group-theoretical
arguments \cite{Boc68}, the phase transition $\beta$-tin$\to${\it
Imma} can be of second order, whereas the phase transition {\it
Imma}$\to$sh has to be of first order. This conclusion was strongly
supported by an ab initio calculation in our previous work
\cite{Gaa04}.

An intermediate phase similar to the {\it Imma} phase, with a
structure more general than the $\beta$-tin and the sh phase was
considered theoretically before \cite{Nee84} with a prediction that
the phase transitions might be accompanied by soft phonon
modes. Since the {\it Imma} structure results from a distortion and
a relative sublattice shift of $\beta$-tin along the $c$-direction,
which corresponds to a soft $\Gamma$ phonon displacement
\cite{Ack01}, it is worthwhile to investigate the pressure
dependence of the phonon-dispersion curves. Especially, it was
speculated that the superconductivity of these metallic
high-pressure phases of silicon might be enhanced by these soft
modes \cite{Cha84,Mig86}. Available experimental and theoretical
data for superconducting properties
\cite{Cha85b,Dra85,Cha86,Ers86a,Ers86b} did not consider the {\it
Imma} phase, because its existence was not known at that time. In
the experiment, the superconducting temperature as a function of
pressure shows a kink \cite{Mig86,Ers86a,Ers86b} within the assumed
stability range of the sh phase. In the corresponding theoretical
work this kink was traced back to some soft modes from calculations
of a few phonon modes under pressure for the sh structure with a
frozen-phonon technique \cite{Dra85}. However, in our opinion this
kink corresponds to the phase transition {\it Imma}$\to$sh. A first
step towards the determination of the superconducting transition
temperature by calculating the electron-phonon coupling is an
analysis of the phonon frequencies of all phases.

In this contribution we examine the phonons of the high-pressure
phases $\beta$-tin, {\it Imma}, and sh and their behaviour near the
phase transitions. In particular, we are interested in soft modes
and the pressure-dependent phonon frequencies. An indication of a
soft-mode behaviour has been already found experimentally: The
low-energy optical phonon frequency of the $\beta$-tin phase at the
$\Gamma$-point decreases with increasing external pressure
\cite{Oli92}. Restricting ourselves to just the $\beta$-tin phase we
did not find a vanishing frequency even beyond the stability range
of this phase \cite{Gaa99, Gaa01}, but we find a lowering of the
lower optical phonon frequency, whereas the degenerate upper ones
increase. In the following we want to include especially the {\it
Imma} and the sh phases in our consideration.

This paper is organised as follows: First, we give a short overview
of the methods on which our calculations are based. Second, we
describe in brief the structures, the relaxation and the pressure
dependence of the structural parameters of the three phases. Third,
phonon-dispersion curves of the three phases are presented as well
as an analysis of the phonon frequencies at the zone-center and at a
zone-boundary point as a function of the external pressure. Finally,
we draw a conclusion.

%
%
The calculations have been carried out using a plane-wave
pseudopotential scheme \cite{Pic70} within the density-functional
theory \cite{Hoh64, Koh65} and the density-functional perturbation
theory \cite{Bar87,Gia91,Gir95} implemented in the PWscf package
\cite{PWSCF}. The ion core of silicon has been described by a
norm-conserving pseudopotential \cite{Ham79,Bac82} created following
a scheme by v.~Barth and Car \cite{Bar}, which is described in
\cite{Cor93}. The exchange-correlation energy is calulated with the
use of the local-density approximation \cite{Per81,Cep80}. The
pressure corresponding to a given structure is obtained by
evaluating the stress tensor \cite{Nie85a,Nie85b}. We have used a
kinetic-energy cutoff $E_{\rm cut}$ of 40~Ry and 1165 special
Monkhorst-Pack points \cite{Mon76} in the irreducible part of the
Brillouin zone with a Methfessel-Paxton \cite{Met89} smearing width
of 0.03~Ry, because of the metallic character of the system. With
these parameters the convergence is better than 0.01~mRy for energy
differences, and the pressure related to the Pulay stress
\cite{Pul69,Pul80} is smaller than 1~kbar. The phonon frequencies
have been calculated on a discrete mesh of 18 points in the
irreduzible Brillouin zone, and the phonon-dispersion curves have
been calculated by Fourier interpolation.
%
%

\begin{figure}[t!]
  \epsfig{figure=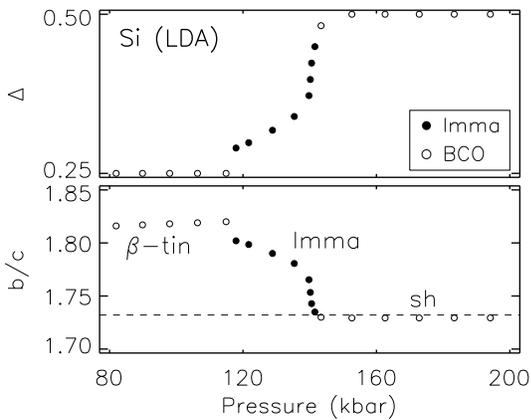,width=7.0cm,angle=0} 
  \caption{Relaxed lattice parameters of the BCO cell as a function
    of the external pressure. The filled symbols mark the stability
    range of the {\it Imma} phase between the $\beta$-tin phase
    ($\Delta=0.25$) and the sh phase ($\Delta=0.5$). The dashed line
    marks the ideal value $b/c=\sqrt{3}$ for the SH
    structure.}\label{PicGS}
\end{figure}

All these three structures have been determined using the same
body-centered orthorhombic (bco) cell with lattice constants
$a\not=b\not=c$ and two atoms at $(0,0,0)$ and $(0,0.5b,\Delta c)$
to which we refer as BCO. In this cell the $\beta$-tin phase can be
reproduced by choosing $a=b$ and $\Delta=0.25$ and the structure of
the sh phase by $b=\sqrt{3}c$ and $\Delta=0.5$. The relaxation of
the lattice parameters has been carried out by evaluating the total
energy as a function of the lattice constants. The equlibrium
lattice parameters for a fixed volume have been determined by the
condition $p_x=p_y=p_z$ where $p_x$, $p_y$, and $p_z$ are the
negative diagonal components of the stress tensor. Therefore, the
external pressure can be taken from the stress tensor since it is
diagonal with equal components for a relaxed structure. For the BCO
structure, the internal parameter $\Delta$ has been relaxed in
addition to the lattice constants, while for the $\beta$-tin and the
sh phases $\Delta$ was fixed at the corresponding value. In the
latter case the ideal relation $a=b$ has been reached for the
$\beta$-tin phase but not exactly for the sh one. On the other hand,
the deviations from $b=\sqrt{3}c$ are negligible and do not affect
the calculation of the phonon frequencies. The results corresponding
to these structures are denoted as BCT and SH.

\begin{figure}[t]
  \epsfig{figure=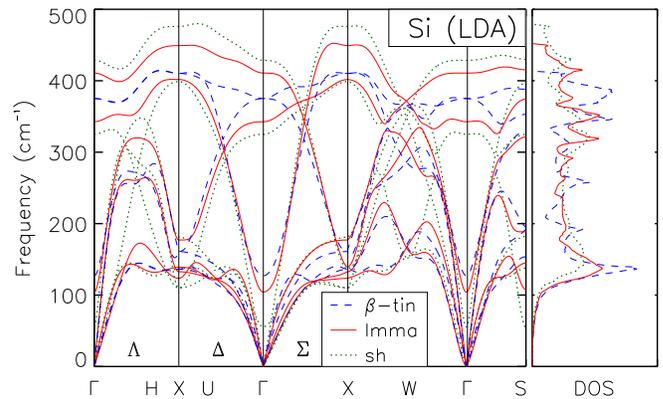,width=8.6cm,angle=0}
  \caption{Phonon-dispersion curves for the $\beta$-tin phase
    ($V=188~a_{\rm B}^3, 98~{\rm kbar}$, dashed), the {\it Imma}
    phase ($V=182~a_{\rm B}^3, 140~{\rm kbar}$, solid), and the sh
    phase ($V=176~a_{\rm B}^3, 183~{\rm kbar}$, dotted) at pressures
    within the stability range of each phase. The corresponding
    directions are from left to right: $\Lambda$ [001], $\Delta$
    [010], $\Sigma$ [100], [111], and [011].
  }\label{PicPhon}
\end{figure}

In fact, for a free relaxation of the BCO cell we obtain the lattice
parameters of the $\beta$-tin phase for large volumes (small
pressures), the ones of the sh phase for small volumes (large
pressures) and between these phases the {\it Imma} phase with
$0.25<\Delta<0.5$ as apparent in Fig.~\ref{PicGS}. Here, with
$\Delta$ being discontinuous the phase transition
$\beta$-tin$\to${\it Imma} seems to be of first order, while the
{\it Imma}$\to$sh transition seems to be of second order because of
a missing discontinuity of $\Delta$. The enthalpy barrier between
the {\it Imma} and the sh phase is only a few meV \cite{Gaa04},
which is less than the precision of our calculations. Therefore, the
determination of the order of the phase transitions from the
behaviour of the lattice parameters as a function of the pressure or
total energy is unreliable. Nevertheless, we can estimate the
transition pressures from Fig.~\ref{PicGS}, and we find that the
$\beta$-tin$\to${\it Imma} transition occurs between 115 and
118~kbar and the {\it Imma}$\to$sh one between 142 and 143~kbar,
which is in good agreement with the experimental values \cite{McM94}
if those data are evaluated in the same way (for further details and
a review of available experimental and theoretical results see
\cite{Gaa04}). In the following we will inspect the phonon spectra
as a possible source for clearer conclusions.

%
%
Phonon-dispersion curves have been calculated for the three phases
at various pressures. For each structure at a given volume within
the stability range of the corresponding phase the phonon
frequencies along selected high-symmetry wave-vector directions are
shown in Fig.~\ref{PicPhon}. Compared to the curves of the
$\beta$-tin phase some degeneracies of freqencies phase at
high-symmetry points are lifted in the {\it Imma} because of the
symmetry of the this phase being lower than that of $\beta$-tin,
see, e.g., at the $\Gamma$- and the X-point. For the sh phase new
degeneracies appear because of the higher symmetry of this phase
with respect to the {\it Imma} phase. Surprisingly, while for the
$\beta$-tin and the {\it Imma} phase the polarisations of the modes
are very similar (even a simultaneous exchange of the polarisation
of two modes along some directions), some major changes appear
between the {\it Imma} and the sh phase. For example, around the
middle of the XU$\Gamma$-direction the third mode (counted from high
to low frequency at U) has a longitudinal acoustic polarisation
whereas it has a transverse optic polarisation in the case of {\it
Imma} and $\beta$-tin.

\begin{figure}[h!]
  \begin{minipage}[b]{17.9cm} 
    \epsfig{figure=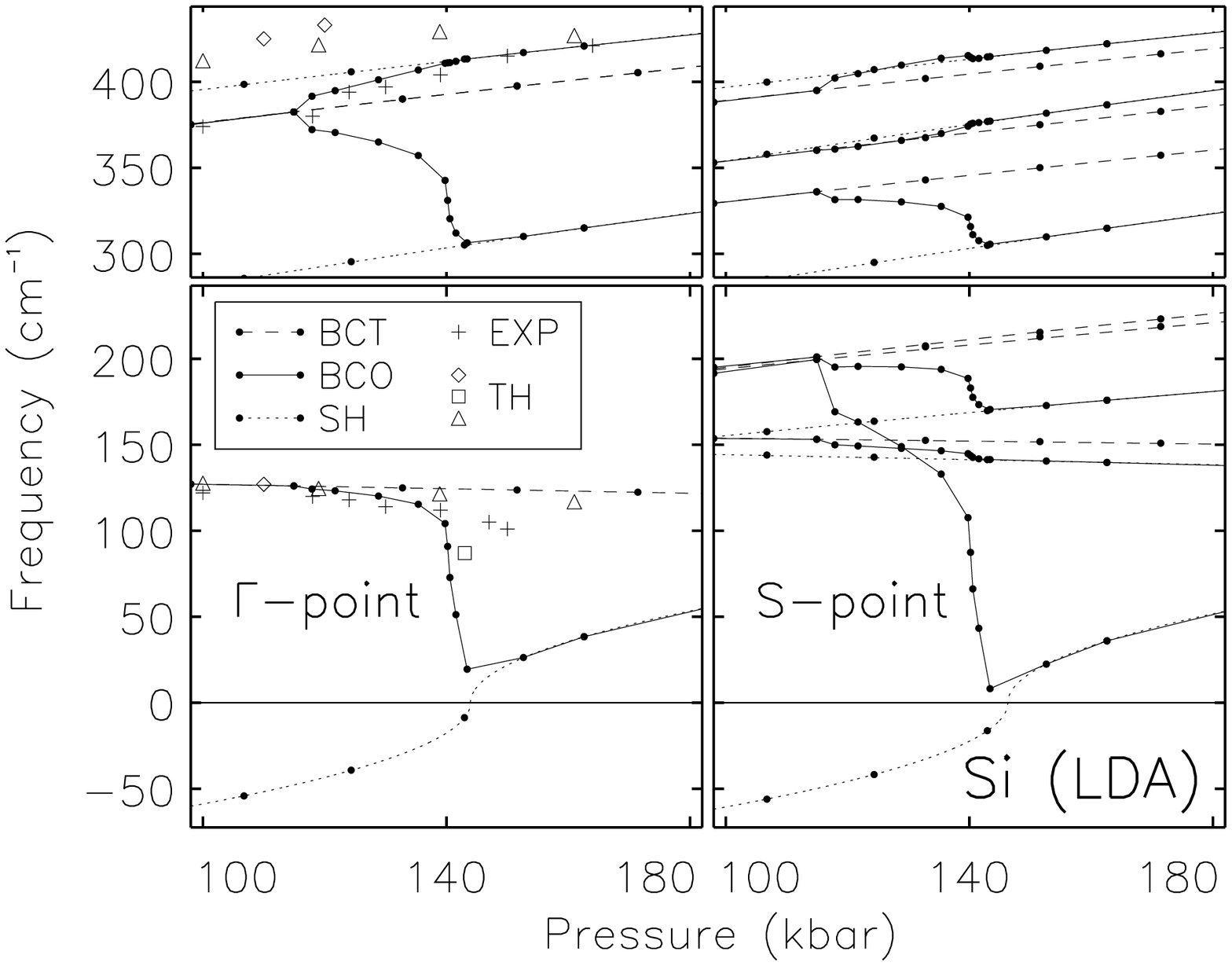,width=8.5cm,angle=0}
    \begin{minipage}[b]{5cm}
      \epsfig{figure=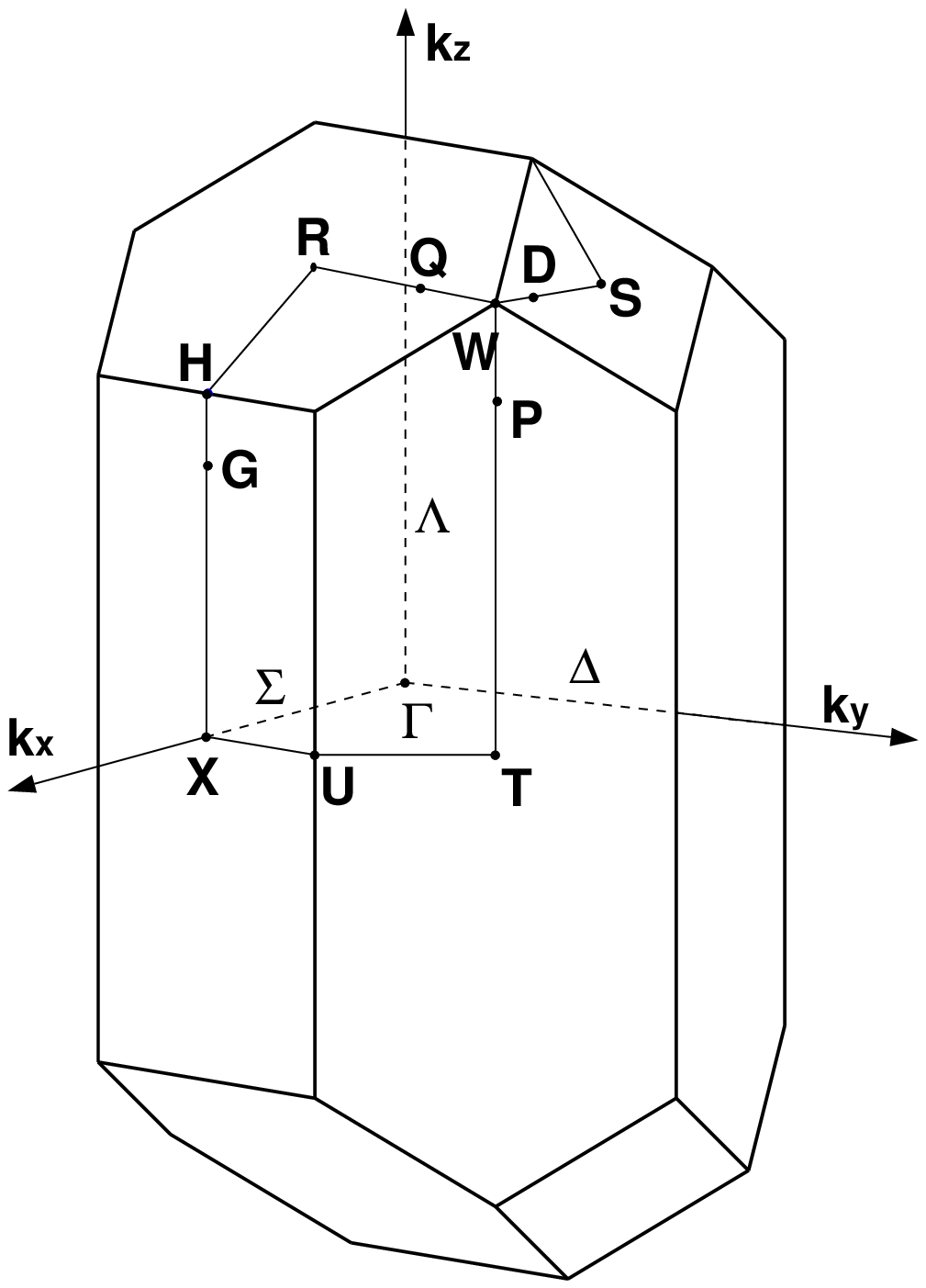, width=3.2cm,angle=0} \\
      \epsfig{figure=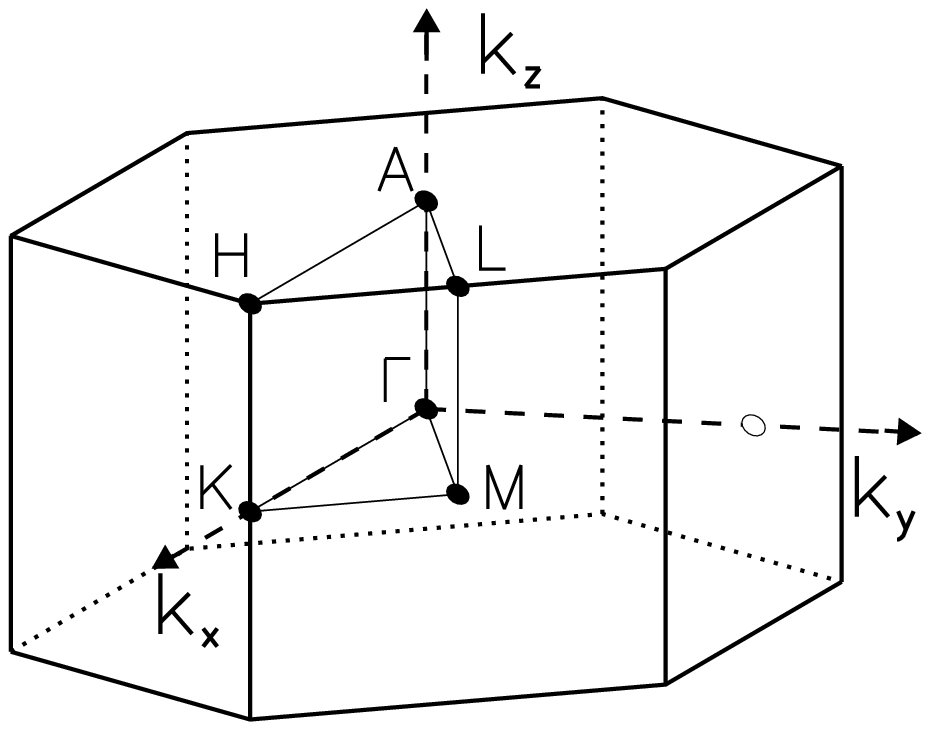, width=2.4cm,angle=0} 
    \end{minipage}
    \caption{Pressure dependence of the phonon frequencies at the
      $\Gamma$- and the S-point for the BCT-, BCO-, and SH
      structure. Imaginary frequencies are drawn along the negative
      frequency-axis. The dots represent calculated values. Lines
      are guides to the eye in the case of BCO, and in the case of
      BCT and SH are derived from the linear function
      $\omega^2(p)$. Reference values from other calculations (TH)
      stem from \cite{Nee84} (square), \cite{Cha85a} (diamonds), and
      \cite{Lew93b} (triangles). Experimental data from \cite{Oli92}
      are denoted as crosses. The pictures of the Brillouin zones
      (bco upper panel, sh lower panel) are used as an illustration.
    }\label{PicGamma}
  \end{minipage}
\end{figure}

Like in the cd phase of Si \cite{Gaa99} the high-energy optic
frequencies increase and the low-energy acoutic ones decrease with
increasing pressure.  In the present case, an exceptionally strong
decrease is found at the $\Gamma$- and the S-points. In order to
find some critical phonon softening we have investigated the
frequencies as a function of external pressure at these points. For
this reason we were not only taking into account the fully relaxed
BCO structure but also the BCT and SH structures. The results are
presented in Fig.~\ref{PicGamma}. Here the three phases are visible
again, meaning that at small pressures the frequencies of BCO are
identical with the ones from the BCT structure corresponding to the
stability of the $\beta$-tin phase. At high pressures the same
correspondence can be stated for sh. The shift of the frequencies
between $\beta$-tin and {\it Imma} is much smaller than between {\it
Imma} and sh which points to a second-order phase transition for
$\beta$-tin$\to${\it Imma} and a first-order phase transition {\it
Imma}$\to$sh. We assume that the change will become more abrupt
between {\it Imma} and sh for more accurate ground-state
calculations which describe the energy barrier between these phases
with more precision. Thus, we expect a discontinuity of the
frequency at this pressure from a more accurate calculation.

In principle, the softening of a phonon mode indicates an
instability against a displacive phase transition.  We find a
softening with decreasing pressure in the SH structure for the sh
phase at the zone-enter $\Gamma$-point and at the zone-boundary
S-point around the sh$\to${\it Imma} phase transition. The
instabilities at both points are related to an instability at the
L-point of the Brillouin zone of a simple-hexagonal cell using the
corresponding monatomic structure. A softening at a point on the
zone boundary corresponds to a doubling of the unit cell. In fact,
in the ideal SH symmetry the BCO unit cell contains two sh unit
cells. We can estimate the transition pressure for the sh$\to${\it
Imma} transition from these soft modes and find the values 143.9 and
146.3~kbar from the softening at the $\Gamma$- and the S-point,
respectively. These transition pressures are in good agreement with
the ones obtained from the relaxed lattice parameters.

\onecolumngrid

\twocolumngrid

We find an excellent agreement also with the available experimental
data of Olijnyk \cite{Oli92} who measured the pressure-dependent
phonon frequencies for the $\beta$-tin structure at a time when the
{\it Imma} phase was yet unknown. Wrongly assuming the $\beta$-tin
phase, theoretical work has not been able to reproduce these data
\cite{Cha85a,Lew93b,Gaa99}. Our results agree completely with those
of Ref.\cite{Oli92} in the low-pressure ($\beta$-tin) and the
high-pressure (sh) range and describe the slope in between ({\it
Imma}). Therefore, Olijnyk was actually observing the phonon
frequencies of all these three phases.

In the {\it Imma} phase the TO modes are split into two branches
which are both Raman active. Since the peak in the experimental
Raman spectrum is very broad and the frequency of the maximum is
just a little bit lower than our calculated value we speculate that
the second peak has a lower intensity and is unresolved from the
high-frequency one. Besides, there might be a second-oder background
in the spectra: If the overtone spectra are dominant in the
$\beta$-tin, Imma, and sh phases as they are in the diamond phase,
then the spectra correlate with the DOS of Fig.~\ref{PicPhon} if the
frequency axis is stretched by a factor of 2. In fact, the
experimental intensity near 500~cm$^{-1}$ has a strong positive
slope and is larger in the $\beta$-tin than in the {\it Imma} or sh
phases.

%
%
In conclusion, we have investigated the phonon dispersion curves for
the $\beta$-tin, {\it Imma}, and sh phases of Silicon. While the
order of a phase transition often is derived from the behaviour of
the structural parameters or of the volume, the order of the
pressure-induced $\beta$-tin$\to${\it Imma}$\to$sh phase transitions
can also be obtained from the pressure dependence of the phonon
frequencies. Here we find a small shift of the frequencies between
the $\beta$-tin and the {\it Imma} phase and a large one between
{\it Imma} and sh pointing to a second-order and first-order phase
transition, respectively. Our calculations are able to reproduce the
experimental data of Olijnyk \cite{Oli92} perfectly. This results in
the insight that in the Raman spectra the {\it Imma} and the sh
phases are visible rather than the assumed $\beta$-tin phase. There
is a splitting of the Raman mode at the $\beta$-tin$\to${\it Imma}
transition as well as a critical softening of the LO mode at
$\Gamma$ and S at the sh$\to${\it Imma} transition which has escaped
experimental observation so far.  Since the pressure-dependent
behavior of these phases is known now, it should be a challenge to
find the lower TO mode of the {\it Imma} and the sh phases and also
the softening of the LO mode in the sh structure experimentally.

%
%
Support by the Heinrich B\"oll Stiftung, Germany, is gratefully
acknowledged. K.G.-N. also like to acknowledge G.~Deinzer, U.~Krey,
and S.~de~Gironcoli for very stimulating and
fruitful discussions and G. Onida for carefully reading the
manuscript. 
%
%

\end{document}